\documentclass[useAMS,usenatbib]{mn2e}
\usepackage{graphicx}
\usepackage{times,epsfig}
\def \deg  {\ifmmode{^{\circ}}\else{$^{\circ}$}\fi}
\newcommand{\ha}{H$\alpha$}

\newcommand{\NII}{[N~{\sc ii}]~6548~\&~6584~\AA}

\newcommand{\oiii}{[O~{\sc iii}]~5007~\AA}

\def\vhel{\ifmmode{V_{{\rm HEL}}}\else{$V_{{\rm HEL}}$}\fi}
\def\vsys{\ifmmode{V_{\rm sys}}\else{$V_{\rm sys}$}\fi}
\def\kms{\ifmmode{~{\rm km\,s}^{-1}}\else{~km~s$^{-1}$}\fi}
\def\vlsr{\ifmmode{v_{\rm lsr}}\else{$v_{\rm lsr}$}\fi}

\title[A long trail behind the planetary nebula HFG1] {A long trail
behind the planetary nebula HFG1 (PK 136+05) and its precataclysmic
binary central star V664 Cas}

\author[P. Boumis et al.] {P. Boumis$^{1}\thanks{e-mail:
ptb@astro.noa.gr}$, J. Meaburn$^{2}$, M. Lloyd$^{2}$, and S. Akras$^{1,3}$
\\$^{1}$Institute of Astronomy \&
Astrophysics, National Observatory of Athens, I. Metaxa \& V. Pavlou,
GR--152 36 P. Penteli, Athens, Greece.\\
$^{2}$Jodrell Bank Centre for Astrophysics, University of Manchester,
Manchester, UK, M13 9PL.\\
$^{3}$Astronomical 
Laboratory, Department of Physics, University of Patras, 
26500 Rio-Patras, Greece. \\}

\date{Accepted 2009 March 16. Received 2009 March 16; in original form
2009 February 11}
\pagerange{\pageref{firstpage}--\pageref{lastpage}} 
\pubyear{2009}

\begin{document}  

\maketitle

\label{firstpage}

\begin{abstract}
 
\noindent 

A deep wide--field image in the light of the \ha\ \& \NII\ emission
lines, of the planetary nebula HFG1 which surrounds the precataclysmic
binary system V664 Cas, has revealed a tail of emission at least
20\arcmin\ long, at a position angle of 316\deg. Evidence is
presented which suggests that this is an $\approx$ 10$^{5}$ y old
trail of shocked material, left behind V664 Cas as it ejects matter
whilst ploughing through its local interstellar media at anywhere
between 29 and 59 \kms\ depending on its distance from the Sun.

\end{abstract}


\section{Introduction}
HFG1 around what later was recognized to be a 13.7 mag precataclysmic
binary star named V664 Cas (\citealt{Bo89}; \citealt{Ac90};
\citealt{Ex05}). This stellar close binary system was found to have a
period of 13.96 h and brightness amplitude of 1 mag. \cite{Ex05} give
its distance D = 310 to 950 pc but there is as yet no parallax
measurement to determine this with more certainty. \cite{Sh04} show
that V664 Cas is a precataclysmic variable with an orbital period of
$\approx$~0.58 days and with the inclination of the rotation axis
along a position angle PA~=~177~$\pm$~5\deg.

The first deep wide--field image of HFG1 in the light of the \ha\ and
\NII\ nebular emission lines has now been obtained which reveals what
appears to be a long trail of emission. This discovery will be
reported here. Such trails have been postulated for stellar systems
ejecting material such as the galactic Luminous Blue variable (LBV)
P~Cygni (\citealt{Me99}; \citealt{Bo06} and refs. cited therein), Mira
AB (\citealt{Ma07}; \citealt{Wa07}) and the Large Magellanic Cloud
LBV, R~143 \citep{Me04}. It is suggested that they form as the stars
plough through their local interstellar media (ISM) during periods of
mass ejection. Several PNe with central binary systems, one star
of which has evolved beyond its Mira~A AGB phase, exhibit bow--shocks
emitting optical emission lines, as they plough through the local ISM
(\citealt{Bo90} and refs. therein) but only the PN, Sh2--188, even
without a binary system, \citep{Wa06} has been shown to have a
similar, though somewhat less spectacular, tail as the one revealed
here for this precataclysmic variable star.  
\section{Observations}
The wide--field image of HFG1 nebula shown in Fig. 1, was obtained
with the 0.3 m Schmidt--Cassegrain telescope at Skinakas Observatory,
Crete, Greece on August 2, 2008. The observations were performed with
a 1024 $\times$ 1024 Thomson CCD which provides a 70\arcmin\ $\times$
70\arcmin\ field of view and an image scale of 4\arcsec\ per
pixel. The HFG1 nebula was observed for 900 s through the \ha\ plus
\NII\ filter and for 180 s with the corresponding continuum
filter. Standard IRAF and STARLINK routines were employed for the
data reduction. All frames were bias subtracted and flat--field
corrected using a series of well exposed twighlight flat--fields. The
continuum image was scaled to the same intensity as the emission line
image by comparing the total star signal (above sky) for a number of
unsaturated stars in both images and multiplying the continuum image
by the appropriate factor. The two images were then aligned by
deriving an astrometric solution using the HST Guide star catalogue
\citep{La99}. Finally the scaled and aligned continuum image was
subtracted from the emission line image to eliminate the confusing
star field and reveal the extent of the nebulosity. Note that the
equatorial coordinates quoted in this work refer to epoch 2000.
\section{Discussion}
The image shown in Fig. 1 reveals extended emission from around the
star (diameter $\sim$ 10\arcmin), with the brightest, arc-like,
sharpest edge to the south--east and with a faint tail extending to
the north--west, about 5\arcmin\ wide and extending at least
20\arcmin. The sharp SE rim is likely to be a bowshock and
the long tail, that is aligned with this feature, is likely to be the
remnant trail behind the PN/star as it moves through the ISM.

\begin{figure*}
\centering
\scalebox{0.90}{\includegraphics{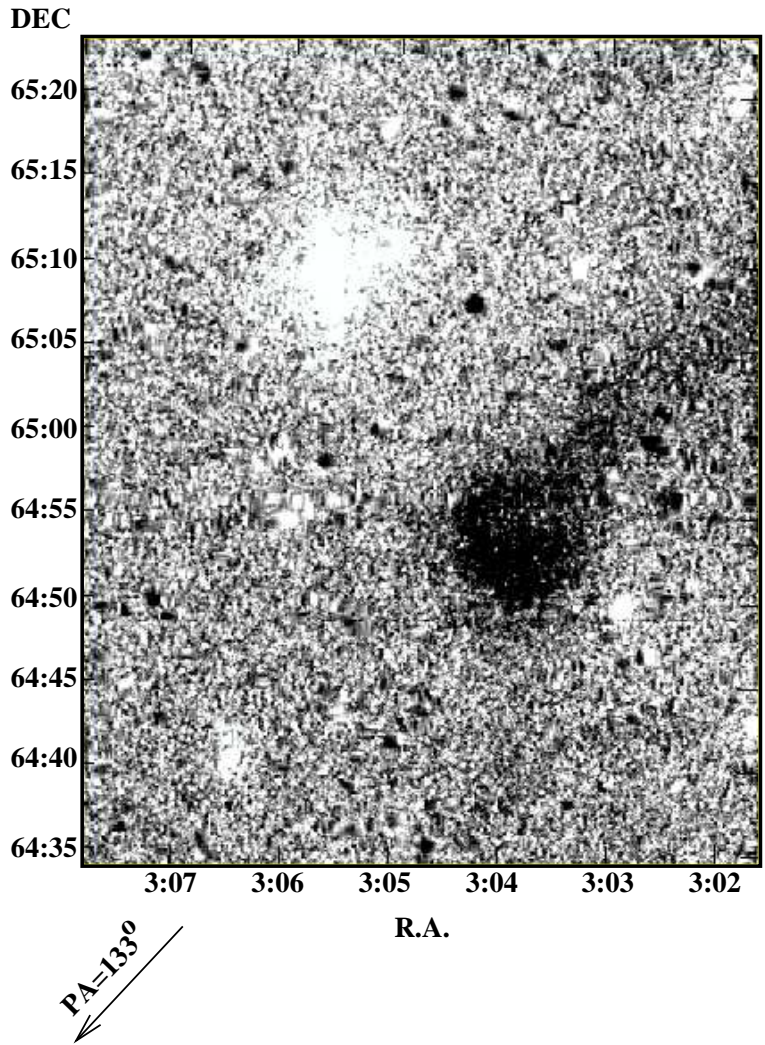}}
\caption[]{A deep, negative grey-scale, representation of the
continuum subtracted \ha\ plus \NII\ image. This is an enlargement
from the wider 70\arcmin\ x 70\arcmin\ field. Only residual star
images are apparent after the subtraction process. The PM of
VV664 Cas is arrowed. This is known to $\pm$ 6\deg\ accuracy.}
\label{fig01}
\end{figure*}

The first test of the suggestion that the extended emission feature
shown in Fig. 1 and directed along PA = 316\deg\ (North through East)
from HFG1 is caused by the passage of the PN through its local ISM is
to determine the proper motion (PM) direction of the progenitor star
V664~Cas, corrected for any relative motion due to differential
galactic rotation. \cite{Ha04} measure the PMs as
d($\alpha$,$\delta$)/dt$=$(7.6,-7.8) mas yr$^{-1}$ where ${\alpha}$ \&
${\delta}$ are the right ascension and declination directions
respectively. These values are accurate to $\pm$ 0.6 mas
yr$^{-1}$. Using Oort's constants A$=$14 and B $=-$12 \kms~kpc$^{-1}$
the correction for differential galactic rotation is $-$2.3 mas
yr$^{-1}$ along galactic longitude, at constant latitude.  The PM
values then become 9.6 $\pm$ 0.6 , -8.9 $\pm$ 0.6 mas yr$^{-1}$. This
correction is independent of distance to V664~Cas. Galactic longitude
at constant latitude increases along PA$=$120\deg\ for V664~Cas. A
PM$=$13 $\pm$ 1.5 mas yr$^{-1}$ along PA$=$133 $\pm$ 6\deg\ due to the
motion of V664~Cas through its local ISM is indicated. This apparent
motion, within the uncertainties, is directly opposite to the
direction of the proposed trail seen in Fig.~\ref{fig01}.

For the range of possible distances to V664 Cas of D$=$310 to 950 pc
as given by \cite{Ex05} the tangential velocity (in the plane of the
sky) is then between 19 $\pm$ 2 and 58 $\pm$ 6 \kms\ for this
corrected PM$=$13 $\pm$ 1.5 mas yr$^{-1}$. The correction for
differential galactic rotation for V664~Cas to its radial velocity for
the same distance range then amounts to -4 to -13 \kms. As the
measured\footnote{\rm http://simbad.u-strasbg.fr/simbad/} local
standard of rest radial velocity of V664~Cas is -26 $\pm$ 2\kms\ then
the radial velocity of HFG1 with respect to its local ISM is between
-22 $\pm$ 2 to -13 $\pm$ 1.3 \kms\ for this same range of possible
distances. Similarly, the range of space velocities again with respect
to its local ISM is therefore V$_{o} =$ 29 $\pm$ 4 to 59 $\pm$ 9 \kms\
tilted towards the observer by between $\theta =$ 49 $\pm$ 5\deg\ to
13 $\pm$ 3\deg\ respectively.  The higher V$_{o}$, and hence the
further distance, would be favoured if the trail found here has been
caused by shock ionization as the mass ejecting stellar system
V664~Cas ploughs through its local ISM. If this is the case, the
measured PMs and the trail angular length of $>$~20\arcmin\ in
Fig.~\ref{fig01} indicate that this process has occurred over a
kinematic age of $\geq$~10$^{5}$ y i.e. simply the angular length of
the trail divided by the stellar PM. This is far too long for the
trail to be ionized by leakage of Lyman photons during the PN phase of
the primary 0.57 solar mass \citep{Sh04} star of the binary system. In
any case there is tentative evidence in Fig.~\ref{fig01} that a
bow--shock in the PA = 133\deg\ direction precedes the motion which
then favours the shock ionization explanation (seen also in the \oiii\
image of \citealt{He82}, in figs. 1 \& 2).

An alternative possibility is that the trail is material ejected in
one direction, along the orbital axis of the central binary system
which is oriented along PA$=$177~$\pm$~5\deg\ \citep{Sh04}. However,
the trail reported here in Fig.~\ref{fig01} is substantially tilted (a
PA difference of 41\deg) with respect to this direction which
precludes any mono--polar ejection mechanism for its
creation. Detailed kinematical observations of both the tail and the
PN itself are now required to investigate the relationships between
the tail, the main part of the PN and the central binary system.
\section*{Acknowledgments} 
We would like to thank the referee for constructive comments that
have improved the paper considerably. Skinakas Observatory is a
collaborative project of the University of Crete, the Foundation for
Research and Technology-Hellas, and the Max-Planck-Institut f\"{u}r
extraterrestrische Physik. This research has made use of the SIMBAD
database, operated at CDS, Strasbourg, France.
\bibliographystyle{mnras}

\label{lastpage}

\end{document}